# Quantum Mechanical Basis of Vision

**Ramakrishna Chakravarthi[1], A. K. Rajagopal[2,3] and A. R. Usha Devi[3,4]**

[1] *Department of Psychology, New York University, NY 10003, USA*
[2] *Center for Quantum Studies, George Mason University, Fairfax, VA 22030, USA*
[3] *Inspire Institute Inc., McLean, VA 22102, USA*
[4] *Department of Physics, Jnanabharathi Campus, Bangalore University, Bangalore-560 056, India*

**Abstract:** The two striking components of retina, i.e., the light sensitive neural layer in the eye, by which it responds to light are (the three types of) color sensitive Cones and color insensitive Rods (which outnumber the cones 20:1). The interaction between electromagnetic radiation and these photoreceptors (causing transitions between cis- and trans- states of rhodopsin molecules in the latter) offers a prime example of physical processes at the nano-bio interface. After a brief review of the basic facts about vision, we propose a quantum mechanical model (paralleling the Jaynes-Cummings model (JCM) of interaction of light with matter) of early vision describing the interaction of light with the two states of rhodopsin mentioned above. Here we model the early essential steps in vision incorporating, separately, the two well-known features of retinal transduction (converting light to neural signals): small numbers of cones respond to bright light (large number of photons) and large numbers of rods respond to faint light (small number of photons) with an amplification scheme. An outline of the method of solution of these respective models based on quantum density matrix is also indicated. This includes a brief overview of the theory, based on JCM, of signal amplification required for the perception of faint light. We envision this methodology, which brings a novel quantum approach to modeling neural activity, to be a useful paradigm in developing a better understanding of key visual processes than is possible with currently available models that completely ignore quantum effects at the relevant neural level.

**Introduction:**

The biological features of the eye and the associated processes of vision [1] offer a unique opportunity to explore the relevance of quantum mechanical principles in understanding this remarkable system. One of the primary functions of the retina (light sensitive layer in the eye) is transduction – converting light into neural signals, which are then processed further by the brain enabling visual perception. Although, traditionally, transduction has been understood in classical terms, a plausible case can be made for a quantum mechanical explanation. The process involves a basic interaction of light (photons) with the quantum levels of the key molecules – the photopigments (such as rhodopsin) residing in the transducers of the eye, namely, **rods** and **cones** - which further modulate the concentrations of various intracellular molecules (e.g.,cyclic Guanosine Monophosphate (cGMP)) and ions (Na+ and K+) thus determining the electrical state of the receptors.

It is well known that rods detect dim light (small number of photons) but are insensitive to color. Several rods act together to amplify the light into a useful neural signal. At high intensities, they are saturated and do not provide any useful interpretable neural signal. On the other hand, cones detect bright light (tens of hundreds of photons) but cannot react to dim light. Furthermore, there are three types of cones sensitive to, respectively, three different wavelength ranges within the visible light spectrum with different peak sensitivities. The interaction between the outputs of these three kinds of cones forms the basis of color vision. We have here a composite system of interacting photons and at least two types of matter systems – that of rods and cones, respectively. The excitations caused by the interactions are somehow statistically correlated (and processed later) to lead to the formation of a coherent visual image. Thus, it seems clear that one can approach visual perception phenomena from a quantum mechanical point of view.

The tools of quantum mechanics of composite systems [2] involve setting up a suitable, tractable model Hamiltonian describing the basic interactions in the system and express the density matrix of the system in terms of the eigensolutions of the Schrodinger equation. In our case, the density matrix would describe the system of rods or cones, given the initial specification of the transduction process. The effects of interactions and intra- and inter-correlations among the rods or cones are then contained in this density matrix. There are various mathematical principles and techniques to extract the physical information of interest from this density matrix. Possible predictions arising out of such an inquiry may lead to experimental investigations as to the relevance of quantum features in this system. We will now outline the suggested models and procedures in some detail.

**Models of interaction of light with rods and cones:**

We first give a brief outline of an exactly soluble Jaynes-Cummings model (JCM) [3] of interaction of a one-mode of photon (quantized electromagnetic field) with a two-level atom (molecule). Suitable modifications of this model are then suggested for describing the interaction of light with rod and cone systems. The JCM Hamiltonian is

$$H = \omega\left(a^+ a + 1/2\right) + \omega_a/2\, \sigma_z + g\left(\sigma_+ a + \sigma_- a^+\right) \quad (1)$$

The first term in this expression represents the photons with photon (light) frequency $\omega$. The second term represents the two-levels of the atom (molecule) with $\omega_a$, the energy separation between the two levels. The last term is the interaction of light with the two-level system with g, the coupling strength. Here $a, a^+$ represent the destruction and creation operators of the photon, with $a^+ a$ the operator representing the number of photons, $a^+ a |n\rangle = n|n\rangle, n = 0,1,2,\cdots$. Also, $a^+|n\rangle = \sqrt{n+1}|n+1\rangle, a|n\rangle = \sqrt{n}|n-1\rangle$. The atom (molecule) is conveniently represented by the Pauli operators. $\sigma_z = \begin{pmatrix} 1 & 0 \\ 0 & -1 \end{pmatrix}, \ \sigma_+ = \begin{pmatrix} 0 & 1 \\ 0 & 0 \end{pmatrix}, \ \sigma_- = \begin{pmatrix} 0 & 0 \\ 1 & 0 \end{pmatrix}$. If $|e\rangle, |g\rangle$ represent the upper (excited) and lower (ground) states of the atom, then we may express the Pauli operators in the form, $\sigma_z = |e\rangle\langle e| - |g\rangle\langle g|, \ \sigma_+ = |e\rangle\langle g|, \ \sigma_- = |g\rangle\langle e|$. With these definitions, the interaction term in

eq.(1) represents the absorption of one photon involves an accompanied transition from ground state to the excited state and emission of one photon is accompanied by the transition from excited state to ground state. Rods and Cones are two types of Photoreceptors. We will now set up the JCM – type models for these two systems.

**Rods** detect dim light (small number of photons) and several rods act together to amplify light signals. In fact a single photon can evoke detectable electrical response. This gives the clue that in JCM, one need only use two photon states, n=0 and n=1, but a large number, N, of two-state system representing the rods. Rods pool together in the bipolar cell enabling detection of dim light. The bipolars combine signals in a coherent way to obtain an amplified signal, particularly by calculating the difference between the number of rods detecting no photon and those detecting one photon. Rods are not sensitive to color! Perhaps one could use Dicke model [4] for the rods to capture these features.

$$H_{Rods} = \hbar \omega a^+ a + \hbar \omega_a \sum_{i=1}^{N} \sigma_{zi} + g \sum_{i=1}^{N} (\sigma_{+i} a + h.c.) \qquad (2)$$

Here the first term represents the photons in number representation, the second, the two-state system of rods, and the third, the photon - rod interaction with **g** as the interaction strength between the photon and rod, assumed to be the same since the rods are identical. We will use the total spin representation to deal with the N rods as a whole and the Dicke states to represent them. This is a way of taking into account the cumulative effect of dim light acting on the rod system mentioned above.

**Cones** require tens or hundreds of photons to evoke similar response – they respond well to bright light (day light). Bipolar cells receive inputs from single cones, especially in the fovea (center and most sensitive part of the retina) and hence there is hardly any pooling of cone signals. There are three types of cones that are sensitive to the specific bands of wavelengths in the visible spectrum (frequencies of light - 400 to 700nm wavelengths) and have different peak sensitivities – that is, they are primarily responsible for our color vision. The model here is then the JCM with three types of photons and three types of two-level systems but allowing large numbers of photons to interact within the three systems. We propose a model that represents interaction between the three cone systems leading to suitable electrical/neural output – a color image! Here we have a single two-level system involved in large number of photons. The three Cone systems ought to be put together by a model Hamiltonian to get a combined "photon" output.

$$H_{Cones} = \sum_{k=1}^{3} \left( \hbar \omega_k a_k^+ a_k + \hbar \omega_{ak} \sigma_{zk} \right) + g_{12} \left( \sigma_{+1} a_2 + h.c. \right) + (23) + (31) \qquad (3)$$

The first term in the sum represents the three primary peak sensitivities (colors) of photons while the second represents the three types of cones each responding to the light of the corresponding color - the cones are color specific to a large extent. The second set of terms represents small interaction between the photon of one type with the cone sensitive to the other type. Here we have chosen a cyclic set for aesthetic reasons of modeling only. These terms lead to interaction between the cones to simulate the composite output of the system. This could have been modeled differently e.g. direct (Heisenberg) interaction between the cones but the above choice seemed more basic based solely on the interactions already introduced! The three cones interact in tandem and are sensitive to large number of photons unlike the rods that operate at low intensities of light.

The total number of photons involved is fixed, representing the intensity of light received by the eye. The two models involve large number of atoms in case of Rods with small number of photons (dim light) and large numbers of photons in case of three types of Cones, but both models must include aspects of "entanglement" to get reasonable outputs! This is the challenge. All these connect to the ganglion in another JCM-like model proposed earlier by Ramakrishna and Rajagopal [5].

**Outline of the method of solution:** The method of solution involves obtaining the solutions to the Schrodinger equations associated with the Hamiltonians described in eqs. (2 and 3,). This first step involves finding the constants of motion associated with each of these Hamiltonians. They are:

**Hamiltonian** in Eq.(2)

Define the collective spin operators: $s_\alpha = \frac{1}{2}\sum_{i=1}^{N} \sigma_{\alpha i}$, $\alpha = x, y, z$. Then eq.(2) has the form

$$H_{Rods} = \omega a^+ a + \omega_a s_z + g(s_+ a + s_- a^+) \quad (4)$$

The following operator commutes with the Hamiltonian:

$$C = a^+ a + s_z \quad (5)$$

**Hamiltonian in Eq.(3)**

Here the operator combination that commutes with the Hamiltonian given in eq.(4) is

$$C_{Cones} = \sum_{k=1}^{3}(a_k^+ a_k + s_k) = \sum_{k=1}^{3} C_k \quad (6)$$

Equation (4) is just the usual JCM with three types of atoms and three types of radiation but the interaction introduced is the added new term to represent the interaction. The operator combination in eq.(8) is designed to commute with these interaction terms and so we obtain a coupled set of JCM-like equations describing the rods. The solutions to these are being studied.

**Photon number amplification [6]**

Without giving the details of our theory, we will here give the essence of our formalism. We rely on the constant of the motion, eq.(5) in developing this theory. The implication of this is that the sum of the number of photons and the number of atoms in the excited states is a constant under time evolution. This has the natural consequence of swapping the atom-photon numbers resulting in photon number amplification as well as discrimination. Furthermore, it has been recognized for some time that eye is a quantum mechanical measuring device [7] and our theory employs this idea in setting up the theory presented in [6]. Three significant results emerge from such analyses: Threshold time for initial exposure to photons, time of perception (time of maximum detection probability), and discrimination of first few photon states.

The next step is the construction of the density matrices in each of these cases, describing the composite systems of radiation and different types of matter systems. Various techniques of manipulations of the density matrices yield physical quantities of interest such as the distribution of photons in a given rod or cone, correlations between a pair of rods or cones etc. These will be the topics of future work.

**Acknowledgement:** AKR thanks the travel and local support by Indo – US Science and Technology forum. While in India during January – March 2008, Professor Rama Rao and Mrs. Vijayalakshmi in Mysore, and, Mr. Rangarajan and Mrs. Padma in Bangalore supported our work with their generous hospitality enabling AKR and ARU to complete the work reported here. This unusual and uncommon support of basic science cannot go without recording our grateful acknowledgement. It is with great pleasure that AKR thanks Dr. G. M. Borsuk for supporting his work for two decades.